\documentclass{elsart}
\usepackage{hyperref}
\newcounter{bla}
\newenvironment{refnummer}{%
\list{[\arabic{bla}]}%
{\usecounter{bla}%
 \setlength{\itemindent}{0pt}%
 \setlength{\topsep}{0pt}%
 \setlength{\itemsep}{0pt}%
 \setlength{\labelsep}{2pt}%
 \setlength{\listparindent}{0pt}%
 \settowidth{\labelwidth}{[9]}%
 \setlength{\leftmargin}{\labelwidth}%
 \addtolength{\leftmargin}{\labelsep}%
 \setlength{\rightmargin}{0pt}}}
 {\endlist}

\begin{document}
\begin{frontmatter}

\title{f2mma: FORTRAN to {\sl Mathematica} translator}

\author[a]{Andrey S. Siver
}


\address[a]{Institute for High Energy Physics, Protvino, Moscow region, Russia}

\begin{abstract}
	f2mma program can be used for translation a program written in some
subset of the FORTRAN language into {\sl Mathematica} programming language. This subset have been enough to translate GAPP (Global Analysis of Particle Properties) program into {\sl Mathematica} language automatically. A table with Standard Model observables calculated with GAPP({\sl Mathematica}) is presented.

\begin{flushleft}
\end{flushleft}


\end{abstract}

\end{frontmatter}


{\bf PROGRAM SUMMARY}

\begin{small}
\noindent
{\em Manuscript Title:} f2mma: FORTRAN to {\sl Mathematica} translator\\
\\
{\em Authors:} Siver A.S.                                                \\
\\
{\em Program Title:} {f2mma}                                          \\
\\
{\em Journal Reference:}                                      \\
\\  
{\em Catalogue identifier:}                                   \\
\\  
{\em Licensing provisions:} none                                   \\
\\  
{\em Programming language:} Perl                                   \\
\\
{\em Computer:} all                                               \\
\\  
{\em Operating system:} Windows, Unix, ...                                    \\
\\  
{\em RAM:} 2 Mb                                              
\\  
{\em Nature of problem:} \\
	f2mma program allows to translate programs from some subset of FORTRAN programming language to  {\sl Mathematica} system's language (where mathematical objects can be analyzed more efficiently in symbolic form). The general aim of f2mma is translation of GAPP (Global Analysis of Particle Properties) package [1].\\
\\  
{\em Solution method:} \\
	Perl::RecDescent module was used with corresponding FORTRAN grammar and transformation rules.\\
\\
{\em Unusual features:} \\
	f2mma can generate additional new lines in Mathematica output programs.\\
\\
{\em Additional comments:} \\
	(1) f2mma use only some subset of FORTRAN language grammar (for example, f2mma goes not process `goto' operator); 
	(2) f2mma does not make syntax check of input FORTRAN program; 
	(3) translation process can be slow for large FORTRAN programs.\\
\\  
{\em References:}
\begin{refnummer}
\item J.~Erler, hep-ph/0005084          
\end{refnummer}
\end{small}

\href{http://sirius.ihep.su/~siver/f2mma.zip}{DOWNLOAD `f2mma' (1.2 Mb)}

\hspace{1pc}
{\bf LONG WRITE-UP}
\section{Introduction}
In programs written in FORTRAN programming language scientific models of specific field have been usually embedded into codes. It's difficult and impractical to support or to upgrade them especially by non-programmers. So the only aim of the codes is often to carry out numerical calculations.

From other side, it's known that with help of computer algebra system such as {\sl Mathematica} one can represent the scientific model as an object which can be simply analyzed and improved. {\sl Mathematica} provides important capabilities such as:
\begin{itemize}
    \item High-precision calculations;
    \item Special functions;
    \item Equations solvers;
    \item Matrices and vectors manipulations;
    \item Sophisticated programming language;    
    \item 2D and 3D graphics.
\end{itemize}

So, there is an idea to translate FORTRAN codes to {\sl Mathematica} system's language and then carry out some calculation in symbolic form. f2mma program can be used as such translator.

For others motivations see \cite{f2cl}, for example.

A design of f2mma program had not aim to provide tool for translation every FORTRAN program (according to any FORTRAN language standard) into {\sl Mathematica} language but actually to translate GAPP\footnote{GAPP - Global Analysis of Particle Properties}\cite{Erler} package into CAS language  (it's important for us since GAPP contains recent scientific models for electroweak interactions of the particles). This purpose was achieved. In the section \ref{sm-table} we present a table with Standard Model observables calculated with GAPP({\sl Mathematica}).

f2mma is a rather simple program and can be easily extended by user. Special effort was made to document the codes.

\section{An example}
Let's translate a test program (which should print out several 'OK'
messages and expressions equaling to zero) in order to see the basic functionalities of f2mma program: 

\hspace{1pc}
{\bf TEST RUN INPUT}
\bigskip
\begin{verbatim}
c 	--- Test subroutine ---
	subroutine sub1(x)
c	... types of scalar variables are ignored:
	integer x,y
	common /t/ r
	real*8 r
	x=x+1
	return
	end
c	--- Test function f1 ---
	integer function fun(x,y)
	integer x,y
	integer m1(2), m2(2:4), m3(x:y)
c	... test 'data'
	data m1 /11,22/
	print *,'m1(1) - 11 = ', m1(1)-11
	print *,'m1(2) - 22 = ', m1(2)-22
	m3(x)=x+y
	fun = m3(x)
	return
	end
c 	------
c	... '_' symbol is replaced by 'TTT' string:
	integer x, y, z, t, ar_int(1:5,2:5), n, s1, s2, fun
	common /t/ r
	real*8 r1,r2
	double precision dp1, dp2, dp3, dp4, prec
	complex*16 q
	logical flag
	print *, '6 OK messages should appear:'
	x=1
	r1=1.
	r2 = 1 . 3
c	... decimal fraction converts to expressions:
	dp1 = 1/1.d3
	ar_int(1,2) = 1
	dp2= 1 . 123 456 d-7
	flag = .true.
	prec =1.d-100
c	--- Test translation of mathematical expressions: ---
	dp3=-(1+2**x/2)/dsqrt(y**2+3.d0)+1
	if (dp3 -  prec.le. 0) print *,'OK-1'
	z=-x**2
	y=(-x)**3
	if (z-y.eq.0) print *,'OK-2'
	t = z-y
c	--- Complex numbers: ---
	q=(0,1)**2+(1,0)**2
	dp4=cdabs(q)
	if (dp4 - prec .le. 0) print *,'OK-3'
c	--- Test complex expressions with 'if': ---
	if (.true.) then 
		print *, 'OK-4'
	else
	if (.true.) then
        t=1
	else
        t=2
	endif
	endif
	if (t.eq.0 .and. flag.eqv..true.) then
		print *, 'OK-5'
	else if (.true.) then 
	t=4
	else
	t=5
	endif
	n=9
	s1=0
c	--- Test 'do': ---    
	do 10, x=0,n
10	s1=s1+x
	s2=0
	do 20 x=n,0,-1
	s2=s2+x
20    continue
	if (s1.eq.(n*(n+1)/2).and.s1.eq.s2) print *, 'OK-6'
	x=0
	y=0
	r=0.e0
	print *, 'Four 0 on right-hand side should appear:'
	call sub1(n)
	print *, 'n - 10     = ', n-10
	n=fun(1,2)
	print *, 'n - 3      = ', n-3
	end
\end{verbatim}


\hspace{1pc}
{\bf TEST RUN OUTPUT}
\bigskip

\begin{verbatim}
(* c --- Test subroutine --- *)
SetAttributes[sub1,HoldAll];
sub1[x_]:=Module[{ y },
(* c ... types of scalar variables are ignored: *)

(* #common: r *)

x=x+1; 
Return[sub1];
];

(* c --- Test function f1 --- *)

SetAttributes[fun,HoldAll];
fun[x_,y_]:=Module[{ fun,m3,m1,m2 },

 F2MmaDimensions[m1]={{1, 2}}; 
F2MmaDimensions[m2]={{2, 4}}; 
F2MmaDimensions[m3]={{x, y}}; 

 (* c ... test 'data' *)

 (* # Inserting to m1 *)
Module[{F2MmaTemp1, F2MmaTemp2, j, j0, k},
F2MmaTemp1={ 11,22 };
F2MmaTemp2=F2MmaDimensions[m1];
j=Array[j0,Length[F2MmaDimensions[m1]]]; j=Transpose[F2MmaTemp2][[1]];
Do[k=1; m1[Sequence@@j]=F2MmaTemp1[[i]];
	j[[k]]++; 
	While[k<Length[F2MmaTemp2]&&j[[k]]>F2MmaDimensions[m1][[k,2]],
	j[[k]]=F2MmaDimensions[m1][[k,1]];j[[k+1]]++; k++];
	,{i,1,Length[F2MmaTemp1]}];];

 Print["m1(1) - 11 = ",m1[1]-11];

 Print["m1(2) - 22 = ",m1[2]-22];

 m3[x]=x+y; 
 fun=m3[x]; 
 Return[fun];
 ];
 
(* c ------ *)

(* c ... '_' symbol is replaced by 'TTT' string: *)

F2MmaDimensions[arTTTint]={{1, 5},{2, 5}}; 

(* #common: r *)

Print["6 OK messages should appear:"];

x=1; 
r1=1; 
r2=(13*10^(-1)); 
(* c ... decimal fraction converts to expressions: *)

dp1=1/(1*10^(3)); 
arTTTint[1,2]=1; 
dp2=(1123456*10^(-6)*10^(-7)); 
flag=True; 
prec=(1*10^(-100)); 
(* c --- Test translation of mathematical expressions: --- *)

dp3=-(1+2^x/2)/Sqrt[y^2+(3*10^(0))]+1; 
If[ dp3-prec<=0
, Print["OK-1"];
 ];

z=-x^2; 
y=(-x)^3; 
If[ z-y===0
, Print["OK-2"];
 ];

t=z-y; 
(* c --- Complex numbers: --- *)

q=(0+I*(1))^2+(1+I*(0))^2; 
dp4=Abs[q]; 
If[ dp4-prec<=0
, Print["OK-3"];
 ];

(* c --- Test complex expressions with 'if': --- *)

If[ True, Print["OK-4"];
 
, If[ True, t=1;  
, t=2;  ];
 ];

If[ t===0 && flag===True, Print["OK-5"];
 
, If[ True, t=4;  
, t=5;  ];
 ];

n=9; 
s1=0; 
(* c --- Test 'do': ---     *)

For[x=0,x<=n,x+=1,s1=s1+x; ];
s2=0; 
For[x=n,x>=0,x+=-1,s2=s2+x; ];
If[ s1===(n*(n+1)/2) && s1===s2
, Print["OK-6"];
 ];

x=0; 
y=0; 
r=(0*10^(0)); 
Print["Four 0 on right-hand side should appear:"];

sub1[n];
Print["n - 10     = ",n-10];

n=fun[1,2]; 
Print["n - 3      = ",n-3];


\end{verbatim}

\section{Table with Standard Model observables}
\label{sm-table}
In order to check general correctness of f2mma translation we present table with SM observables (that is values of physical expressions arising from Standard Model theory of particle physics). They were calculated with followings values of the fitted parameters: $M_Z=91.187$ {\rm GeV}, $m_t=166.96$ {\rm GeV},
$m_b=4.21$ {\rm GeV}, $m_c=1.29 $ {\rm GeV}, $\alpha_S(0)=0.121$,
$m_H=\exp(4.7126)$ {\rm GeV}, $T=U=S=B=0$,  $Z=3$,  $M_{Z'}=1000$
GeV, $sin\theta=0$, $\lambda_g=1$. Total $\chi^2$ (with parameters
defined above) is about 49.6 for 37 effective degrees of freedom (it's quite comparable with GAPP(FORTRAN) result 49.4 at the minimum of $\chi^2$).

For explanation of the observables abbreviations please see \cite{Erler}, \cite{PDG}.
\newpage
\begin{table}
\caption{Table of the results obtained with GAPP({\sl
Mathematica}). `SM value' -- calculated value, `Exp.' -- experimental value of observable, `Exp. unc.' -- experimental uncertaincy,  `Pull' =  (`SM value' - `Exp.')/`Exp. unc.'}

\begin{tabular}{|l|l|l|l|l|l|} \hline ID & Observable & SM value & Exp. & Exp. unc. & Pull
\\\hline
1 & ${M_Z}$ & 91.187 & 91.1876 & 0.0021 & -0.285714 \\
2 & ${{\Gamma }_Z}$ & 2497.06 & 2495.2 & 2.3 & 0.807306 \\
3 & ${{\sigma }_{{had}}}$ & 41.4737 & 41.541 & 0.037 & -1.81758 \\
4 & ${R_e}$ & 20.7482 & 20.804 & 0.05 & -1.11679 \\
5 & ${R_{\mu }}$ & 20.7483 & 20.785 & 0.033 & -1.11132 \\
6 & ${R_{\tau }}$ & 20.7936 & 20.764 & 0.045 & 0.65709 \\
7 & ${A^{{FB}}}(e)$ & 0.0162736 & 0.0145 & 0.0025 & 0.709421 \\
8 & ${A^{{FB}}}(mu)$ & 0.0162736 & 0.0169 & 0.0013 & -0.481883 \\
9 & ${A^{{FB}}}(tau)$ & 0.0162736 & 0.0188 & 0.0017 & -1.48615 \\
10 & $P(tau)$ & 0.147303 & 0.1439 & 0.0043 & 0.791313 \\
11 & ${P^{{FB}}}(tau)$ & 0.147303 & 0.1498 & 0.0049 & -0.509664 \\
12 & ${{{sin}}^2}({{{{\theta }_e}}^{{eff}}}) ({Q_{{FB}}})$ & 0.0422812 & 0.0403 & 0.0026 & 0.762011 \\
13 & ${A^{{FB}}}(s) (DELPHI+OPAL)$ & 0.10337 & 0.0976 & 0.0114 & 0.506097 \\
14 & ${R_{d,s}}/({R_d}+{R_u}+{R_s})$ & 0.359181 & 0.371 & 0.023 & -0.513853 \\
15 & ${R_b}$ & 0.215635 & 0.21638 & 0.00066 & -1.12942 \\
16 & ${R_c}$ & 0.172329 & 0.172 & 0.003 & 0.109588 \\
17 & ${A^{{FB}}}(b)$ & 0.103262 & 0.0997 & 0.0016 & 2.22614 \\
18 & ${A^{{FB}}}(c)$ & 0.0737844 & 0.0706 & 0.0035 & 0.909841 \\
19 & ${{{A_{{LR}}}}^{{FB}}}(b)$ & 0.934691 & 0.925 & 0.02 & 0.48454 \\
20 & ${{{A_{{LR}}}}^{{FB}}}(c)$ & 0.667872 & 0.67 & 0.026 & -0.0818604 \\
21 & ${A_{{LR}}}(hadrons)$ & 0.147303 & 0.15138 & 0.00216 & -1.88766 \\
22 & ${A_{{LR}}}(leptons)$ & 0.147303 & 0.1544 & 0.006 & -1.18289 \\
23 & ${{{A_{{LR}}}}^{{FB}}}(mu)$ & 0.147303 & 0.142 & 0.015 & 0.35351 \\
24 & ${{{A_{{LR}}}}^{{FB}}}(tau)$ & 0.147303 & 0.136 & 0.015 & 0.75351 \\
25 & ${A_e} ({Q_{{LR}}})$ & 0.147303 & 0.162 & 0.043 & -0.341799 \\
26 & ${{{A_{{LR}}}}^{{FB}}} (s)$ & 0.935665 & 0.895 & 0.091 & 0.446873 \\
27 & ${M_W}(LEP)$ & 80.391 & 80.412 & 0.042 & -0.498945 \\
29 & ${M_W}(Tevatron)$ & 80.391 & 80.45 & 0.058 & -1.01648 \\
30 & ${{\Gamma }_W}(Tevatron)$ & 2.09364 & 2.103 & 0.106 & -0.0883258 \\

\end{tabular}
\end{table}
\newpage
\begin{table}
\caption{Continue of the table with results obtained with
GAPP({\sl Mathematica}).}
\begin{tabular}{|l|l|l|l|l|l|}

31 & ${m_t}(pole)(1, CDF\;I)$ & 177.007 & 176.1 & 7.36 & 0.123296 \\
32 & ${m_t}(pole)(2, CDF\;I)$ & 177.007 & 167.4 & 11.39 & 0.843499 \\
33 & ${m_t}(pole)(3, CDF\;I)$ & 177.007 & 186. & 11.51 & -0.781281 \\
34 & ${m_t}(pole)(1, D0\;I)$ & 177.007 & 180.1 & 5.34 & -0.579128 \\
35 & ${m_t}(pole)(2, D0\;I)$ & 177.007 & 168.4 & 12.84 & 0.670363 \\
36 & ${m_t}(pole)(1, CDF\;II)$ & 177.007 & 177.5 & 13.15 & -0.0374557 \\
37 & ${m_t}(pole)(2, CDF\;II)$ & 177.007 & 175. & 18.95 & 0.105934 \\
43 & ${m_c}({m_c})$ & 1.48297 & 1.46484 & 0.17046 & 0.106346 \\

44 & ${m_b}({m_b})$ & 1.96517 & 1.97361 & 0.162013 & -0.0521324 \\
45 & ${{{{{\Delta \alpha }}_{{had}}}}^3} \rm (1.8GeV)$ & 0.0058189 & 0.005768 & 0.0001 & 0.509 \\
46 & $({g_{\mu }}-2-\alpha/\pi)/2$ & 4509.3 & 4511.07 & 0.8 & -2.2071 \\
47 & ${R_{\tau }}$ & 292.3 & 290.87 & 0.52 & 2.75057 \\
48 & ${{{g_L}}^2}(NuTeV\; 2002)$ & 0.303984 & 0.30005 & 0.00137 & 2.87117 \\
49 & ${{{g_L}}^2}(NuTeV\; 2002)$ & 0.0300695 & 0.03076 & 0.0011 & -0.627763 \\
50 & $kappa(CCFR\; 1997)$ & 0.583381 & 0.582 & 0.0041 & 0.336865 \\
51 & ${R_{\nu }}(CHARM\; 1984)$ & 0.309273 & 0.3021 & 0.0041 & 1.74949 \\
52 & ${R_{\nu }}(CDHS\; 1984)$ & 0.309273 & 0.3096 & 0.0043 & -0.0760666 \\
53 & ${R_{\overline{\nu }}}(CHARM\; 1984)$ & 0.386309 & 0.403 & 0.016 & -1.0432 \\
54 & ${R_{\overline{\nu }}}(CDHS\; 1984)$ & 0.386309 & 0.384 & 0.018 & 0.128266 \\
55 & ${R_{\overline{\nu }}}(CDHS\; 1979)$ & 0.381726 & 0.365 & 0.016 & 1.0454 \\
56 & ${{{g_V}}^{(\nu ,e)\: }}(CHARM\;II)$ & -0.0392145 & -0.04 & 0.015 & 0.0523635 \\
57 & ${{{g_A}}^{(\nu ,e)\: }}(CHARM\;II)$ & -0.506548 & -0.507 & 0.014 & 0.0323025 \\
58 & ${g_{{ee}}}(SLAC\;E158)$ & 0.0441208 & 0.059 & 0.0122 & -1.2196 \\
60 & ${Q_W}(Cs)$ & -73.2353 & -72.84 & 0.46 & -0.859329 \\
61 & ${Q_W}(Tl)$ & -116.879 & -116.4 & 3.64 & -0.131507 \\
62 & $\ln B(b\rightarrow s \gamma )/B(b\rightarrow c  e \nu )$ & -5.73228 & -5.69 & 0.17 & -0.248685 \\
63 & $A^{FB}(e)(CDF\,{II})$ & 0.231486 & 0.2238 & 0.005 & 1.53724
\\\hline
 \end{tabular}
\end{table}
\newpage

\section{Acknowledgments}
Author would like to thank Zenin O. V., Ezhela V. V., some Perl
experts from public forum http://forum.vingrad.ru.

The work was not supported by the project RFFI-05-07-90191-w.

\end{document}